\documentclass[conference, onecolumn]{IEEEtran}
\usepackage{soul} 
\usepackage{placeins} 
\usepackage{longtable} 

\usepackage[english]{babel}

\ifCLASSINFOpdf
\else
\fi
\ifCLASSINFOpdf
   \usepackage[pdftex]{graphicx}
\else
\fi

\usepackage{float}
\usepackage{placeins}
\usepackage{graphicx}
\usepackage{color}
\usepackage{caption}
\usepackage{pbox}
\usepackage{rotating}
\usepackage{adjustbox}
\usepackage{dcolumn}
\usepackage{enumerate}
\usepackage{amssymb}
\usepackage{float}
\usepackage{epstopdf}
\usepackage{amssymb}
\usepackage{amsthm}
\usepackage{mathtools}
\usepackage{caption}
\usepackage{subcaption}
\usepackage{tablefootnote}
\usepackage{amsthm}
\usepackage{amsmath,amssymb}

\usepackage{longtable} 
\usepackage{color}
\usepackage{fancyhdr}

\usepackage{booktabs}
\usepackage{xtab,afterpage}
\usepackage{tablefootnote}
\usepackage{csquotes}

\usepackage{hyperref}
\renewcommand{\thispagestyle}[2]{} 

\fancypagestyle{plain}{
        \fancyhead{}
        \fancyhead[C]{first page center header}
        \fancyfoot{}
        \fancyfoot[C]{first page center footer}
}
\begin{document}

%

\title{Bull Bear Balance: A Cluster Analysis of Socially Informed Financial Volatility}


\author{\IEEEauthorblockN{Jonathan Manfield}
\IEEEauthorblockA{Computer Science Department\\
University College London (UCL)\\
\texttt{j.manfield@cs.ucl.ac.uk}}
\and
\IEEEauthorblockN{Derek Lukacsko}
\IEEEauthorblockA{Computer Science Department\\
University College London (UCL) \\
\texttt{derek.lukacsko.15@ucl.ac.uk}}
\and
\IEEEauthorblockN{Th\'arsis T. P. Souza}
\IEEEauthorblockA{Computer Science Department\\
University College London (UCL)\\
\texttt{t.souza@cs.ucl.ac.uk}}}


%


\maketitle

\begin{abstract}
Using a method rooted in information theory, we present results that have identified a large set of stocks for which social media can be informative 
regarding financial volatility. 
By clustering stocks based on the joint feature sets of social and financial variables, 
our research provides an important contribution by characterizing the conditions in which social media signals can lead financial volatility. 
The results indicate that social media is most informative about financial market volatility when the ratio of bullish to bearish sentiment is high, 
even when the number of messages is low. 
The robustness of these findings is verified across 500 stocks from both NYSE and NASDAQ exchanges. 
The reported results are reproducible via an open-source library for social-financial analysis made freely available.

\end{abstract}



%
\IEEEpeerreviewmaketitle

\section{Introduction}
Recent research has dampened the consensus about the unpredictability of market prices. 
As a result, predicting the stock market has garnered the attention of diverse fields and domains where different information channels have been explored, 
such as news~\cite{tetlock2007giving, tetlock2008more, Tobias:2013}, search engines~\cite{PreisCurme:2014,Preis5707, citeulike:12299800, JOFI:JOFI1679} 
and, more recently, social media~\cite{Bollen20111, citeulike:13108056, 1507.00784, 2016arXiv160104535S}.

Social media is of particular interest due to the high volume and velocity of activity of an ever-evolving network that is constantly providing and creating information. 
It offers real-time information that can be attributed to specific people, events, markets, and securities.
This is channeled through the use of so-called \textit{cashtags} (e.g., `\$AAPL') in messages, enabling the creation of feeds associated with particular stock symbols. 
The use of \textit{cashtags} has since been propagated onto Twitter, thus providing a means to direct social sentiment to the specific stock or 
securities of that it references. 
As a result, social media user sentiment can be harnessed as a source of information with respect to financial market activity.

Now, the opinions of traders, professional bloggers and analysts, along with laypeople's opinions, 
are aggregated in a dynamic social network that can possibly explain some of the variance in market behavior. 
Initial research~\cite{yahoo1} sought to quantify a relationship between social media analytics with financial market data such as daily returns. 
The observed results outperformed baseline trading strategies, providing evidence that the volume of tweets can reduce uncertainty about financial returns. 
However, volume-based methods disregard any possible predictive information from \textit{qualitative} aspects of the data (i.e., the actual content or content polarity). 
By applying sentiment analysis techniques \cite{1507.00955} to a corpus of tweets, 
a sentiment score or emotion classification can be derived to quantify this qualitative dimension~\cite{1507.00955}. 
The effectiveness of this semantic approach has been examined in~\cite{Bollen20111}, where collective moods analyzed from large 
collections of daily tweets were used to increase an existing financial predictor's performance to an accuracy of 87.6\%.

While a broad analysis across several financial securities might reveal that social signs are relevant for explaining financial dynamics to some extent~\cite{Bollen20111, citeulike:13108056},
little is known about confounding factors that distinguish assets with predictive social signs from assets with no extra information provided by social media.
Leveraged by a non-parametric analysis founded in information-theoretic measures, we demonstrate that social signs can be useful for most stocks from both NYSE and NASDAQ exchanges.
This alone is an interesting and very sound result, compared to current literature' however, we extend this analysis to provide, for the first time,
possible explanations of features that might be essential to distinguish predictive social signs from non-predictive ones.

\subsection{Research Questions}

\begin{itemize}
    \item \textbf{RQ1. \textit{Which stocks from the NASDAQ and NYSE exchanges exhibit a significant information surplus when using social media as a leading indicator of the stocks' future volatility?}} \\ 
    For each stock, lead-lag mutual information analysis is performed between financial and social media data.


    \item \textbf{RQ2. \textit{Under what configuration of social media and financial variables are social media analytics informative of future financial movements?}} \\
    We aim to determine the feature profile (using financial and social media variables) of companies that are most indicative of a statistically significant 
    social media lead-time information.
\end{itemize}


\section{Methodology}

\subsection{Data}
The data correspond to the daily social media analytics and market quotes of NYSE and NASDAQ listed companies between 01-01-2012 and 01-01-2016.

Social media data were provided by PsychSignal \cite{PsychSignal}, which operates a customized 
Twitter and StockTwits collection framework tracking messages containing \textit{cashtags} (e.g., \$AAPL). 
State-of-the-art natural language processing algorithms were applied to relevant messages, 
labeling each with a sentiment disposition (i.e., bullish or bearish) and a measure of disposition intensity. 
A daily aggregate of these data was provided for each tracked stock. 
Furthermore, for each stock with available social media data, we also considered the historic records of daily stock quotes.
The following data sets were utilized:
\begin{itemize}
    \item \textbf{PsychSignal Social Media Database.} Sentiment daily aggregates containing the following stock quotes attributes:
\begin{itemize}
  \item \texttt{SYMBOL}: the stock symbol (ticker) to which the sentiment data refers to;
  \item \texttt{TIMESTAMP\_UTC}: the date and time of the analyzed data in UTC format;
  \item \texttt{BULLISH\_INTENSITY}: positive sentiment polarity;
  \item \texttt{BEARISH\_INTENSITY}: negative sentiment polarity;
  \item \texttt{BULL\_SCORED\_} \texttt{MESSAGES}: positive sentiment volume, number of messages;
  \item \texttt{BEAR\_SCORED\_} \texttt{MESSAGES}: negative sentiment volume, number of messages;
  \item \texttt{TOTAL\_SCANNED\_} \texttt{MESSAGES}: total number of scanned messages.
\end{itemize}
    \item \textbf{Google Finance.} Daily Market Quotes:
    \begin{itemize}
    \item \texttt{OPEN}: daily opening price;
    \item \texttt{HIGH}: daily high price;
    \item \texttt{LOW}: daily low price;
    \item \texttt{CLOSE}: daily closing price;
    \item \texttt{VOLUME}: financial volume, number of daily trades.
    \end{itemize}
\end{itemize}

\subsubsection{Variables Analyzed}
We used a value of Daily True Range (TR) (see Equation \ref{TRequation}) as a measure of financial volatility. 
A time series of volatility data for each stock was derived using financial quotes as follows:
\begin{equation}\label{TRequation}
    \begin{split}
        TR_{t} = \textup{max}[(\textup{High}_{t}-\textup{Low}_{t}), \\ (\textup{Low}_{t}-\textup{Close}_{t-1}),\\(\textup{High}_{t}-\textup{Close}_{t-1})]
    \end{split}
\end{equation}

We applied a principal component analysis (PCA) to a set of social media variables in order to obtain a time series that contains the majority of underlying information from the sentiment features considered.
\texttt{PCA\_SOCIAL\_CHANGE} is defined as the daily change in the time series obtained by applying the PCA and extracting 
the first principal component from the set of social media variables considered, as defined in Table \ref{sentvariables}.
Principal component analysis is a dimensionality reduction technique that we utilized for feature extraction. 
The PCA permits the reduction of numerous correlated, co-linear variables to a component (or feature set of components). 
Tables \ref{sentvariables} and \ref{finvariables} list the features utilized in this work. 
\begin{table}[h]
    \centering
    \caption{Social Media Features}
    \label{sentvariables}
    \begin{tabular}{llp{6cm}}
    \toprule
     & \textbf{Feature} & \textbf{Description} \\ \midrule
    1 & \texttt{BULLISH\_INTENSITY} & positive sentiment polarity\\
    2 & \texttt{BEARISH\_INTENSITY} & negative sentiment polarity \\
    3 & \texttt{BULL\_MINUS\_BEAR} & the ratio of 1 to 2 \\
    4 & \texttt{BULL\_SCORED\_} \texttt{MESSAGES} & positive sentiment volume, number of messages\\
    5 & \texttt{BEAR\_SCORED\_} \texttt{MESSAGES} & negative sentiment volume, number of messages \\
    6 & \texttt{BULL\_BEAR\_MSG\_} \texttt{RATIO} & volume of bullish messages over volume of bearish messages \\
    7 & \texttt{TOTAL\_SCANNED\_} \texttt{MESSAGES} & total messages, including neutral sentiment \\
    8 & \texttt{LOG\_BULL\_RETURN} & log difference in daily volume of bullish messages \\
    9 & \texttt{LOG\_BEAR\_RETURN} & log difference in daily volume of bearish messages \\
    10 & \texttt{LOG\_BULLISHNESS} & log difference between 4 and 5 \\
    11 & \texttt{LOG\_BULL\_BEAR\_} \texttt{RATIO} & log ratio between 4 and 5 \\
    12 & \texttt{LOG\_BULL\_MINUS\_} \texttt{BEAR\_CHANGE} & log daily difference in 3 \\ 
    13 & \texttt{TOTAL\_SCANNED\_} \texttt{MESSAGES\_DIFF} & daily difference in 7 \\
    14 & \texttt{TOTAL\_SENTIMENT\_} \texttt{MESSAGES\_DIFF} & daily difference in volume for messages with non-neutral polarity \\
    15 & \texttt{PCA\_SOCIAL\_} \texttt{CHANGE} & First principal component derived from 8, 9, 10, 11, 12, 14, and 15
    \\ 
    \bottomrule
    \end{tabular}
\end{table}

\begin{table}[h]
    \centering
    \caption{Financial Features}
    \label{finvariables}
    \begin{tabular}{llp{6cm}}
    \toprule
     & \textbf{Feature} & \textbf{Description} \\ \midrule
    16 & \texttt{OPEN} & daily opening price\\
    17 & \texttt{HIGH} & daily high price \\
    18 & \texttt{LOW} & daily low price\\
    19 & \texttt{CLOSE} & daily closing price\\
    20 & \texttt{VOLUME} & financial volume, number of daily trades \\
    21 & \texttt{LOG\_RETURN} & percent change in log close price \\
    22 & \texttt{LOG\_CLOSE} & log closing price\\
    23 & \texttt{LOG\_HIGH} & log daily high price \\
    24 & \texttt{LOG\_LOW} & log daily low price \\
    25 & \texttt{VOLATILITY\_1} & the absolute value of the difference between 22 and 24 \\
    26 & \texttt{VOLATILITY\_2} & the absolute value of the difference between 22 and the previous day's 21 \\ 
    27 & \texttt{VOLATILITY\_3} & the absolute value of the daily difference between 24 and the previous day's 21 \\
    28 & \texttt{TR} & the max between 25, 26, and 27 \\
    29 & \texttt{LOG\_VOLUME\_DIFF} & log daily difference in 20 \\
    30 & \texttt{LOG\_TR\_DIFF} & log daily difference in 28 \\
    \bottomrule
    \end{tabular}
\end{table}


\subsection{Information Surplus}
Information surplus~\cite{citeulike:13108056} was derived from mutual information, a measure of the mutual dependence between two data sets. 
Let $S$ be the variable \texttt{PCA\_SOCIAL\_CHANGE}, which is the daily change in the time series obtained by applying the PCA 
and extracting the first principal component from the set of social media variables as defined in Table \ref{sentvariables} and 
let $F$ be \texttt{LOG\_TR\_DIFF}, which is the log of the daily difference in the Daily True Range (see Equation \ref{TRequation}).
If the addition of series $S$ provides information about the movements of series $F$, then it is said that 
a dependency or mutual information ($MI$) exists between $S$ and $F$. However, such a dependency is non-directional; 
in order to determine whether $S$ \textit{leads} $F$, $S$ must provide more information on a lagged series of $F$ than the baseline $MI$ (i.e. non-lagged). 
Determining whether a baseline dependency exists between the series of a social media feature $S_{l=0}$ and a 
financial feature $F_{l=0}$ on the same day is the first step in identifying if $S$ leads or is predictive of $F$. 
Mutual information tells us how much the information about $S$ reduces the uncertainty of $F$. 
Equation (\ref{MI}) shows the formal form, where we attempt to reduce uncertainty or 
increase information by taking the double integral over the log of the joint entropy for both series over each distribution. 

\begin{equation}\label{MI}
MI(S;F)=\int\int f(s,\textup{f})\log\frac{f(s,\textup{f})}{f_{s}(s)f_{\textup{f}}(\textup{f})}\,ds\,df 
\end{equation}

The data need to be grouped into bins to determine the mutual information between the two series. This is necessary 
to calculate entropy because the probability of observing an instance $i$ in each bin $s$ and $f$ constitutes the probability distributions. 
The number of bins is dependent on the size of the data; therefore, our bin sizes were typically identical 
per feature (i.e., there are roughly 365 daily instances of tweets and financial data per security per year). 
The bin size $k$ was calculated using Sturge's Rule (see Equation \ref{MI2}), which has been found to be more accurate than comparable methods when used to calculate entropy in the $MI$ algorithm~\cite{bin}.

\begin{equation}\label{MI2}
k = \textup{log}_{2}n + 1
\end{equation}

Mutual information was then computed at consecutive daily time lags. The information \textit{gain} at time lag $l=i$ is 
calculated by finding the difference between mutual information at $l=i$ and $l=0$, where $i$ is a certain day lag 
and $l=0$ corresponds to the baseline case. Information surplus is expressed as a percent of the $MI$ above what we would expect over the given time frame. 
Therefore, if we achieve a surplus above the average $MI$ from $l=i$ to $l=0$, then the social media time series $S$ is said to lead $F$.

\begin{equation}\label{IS}
\textup{Information Surplus}_{l} = \frac{MI(S;F)_{l=i}-MI(S;F)_{l=0}}{MI(S;F)_{l=0}} \times 100
\end{equation}

\subsection{Validating Significant Information Surplus}
To validate whether the information surplus is statistically significant, we verified whether information is more leading than trailing, and we compared the
obtained results with those obtained via a randomly permuted time-series.

We first filtered companies whose surplus is more \textit{trailing} than it is \textit{leading} 
by identifying stocks for which the daily changes in $S$ carry more information about the daily $F$ in hindsight ($l=-i$) 
rather than during the same ($l=0$) or a previous day ($l=i$). 
We calculated $MI$ on a forward shift (\textit{ex-post}) and on a backward shift (\textit{ex-ante}), 
eliminating stocks where the \textit{ex-post} $MI$ is greater than the \textit{ex-ante} $MI$. 
Where the $MI$ for a lag is less than the $MI$ for a retrospective advance of the series (i.e. $MI_{l=i}<MI_{l=-i}$), 
we can assert that $MI(S;F)$ trails more trailing than it leads, and it is thus insignificant.

A random permutation of the remaining symbols was then performed 100 times. With $\alpha = 0.05$, 
the stocks must outperform 95\% of the randomly permuted data to be considered to have a significant surplus. 

\subsection{Clustering}
To identify the conditions under which social signs may be predictive of financial volatility, 
we utilized a clustering method to determine the configuration of social media and financial variables that is 
indicative of a high information surplus. Each stock was represented as a vector of social media and financial variables. 
In addition, the respective scores of information surplus and the features that describe its nature 
(e.g., the size of the lag it was obtained at) were included. Stocks with similar configurations of social media and financial variables were
grouped together in clusters. 

Inspecting the feature profiles of the clusters (i.e., the average representation of constituent stocks' features) 
allows us to identify the social signs of financial dynamics that are indicative of a high information surplus. 
Clusters with a profile containing many significant lags and high overall information surplus will provide us insight into the 
configuration of variables that characterizes a stock with predictable volatility.

We used k-means clustering, which is an unsupervised learning algorithm that partitions instances into \textit{k} clusters by minimizing the within-cluster sum of square error (WCSS) between instances in each set $S$ using a distance metric (see Equation \ref{kmeans}). 

\begin{equation}\label{kmeans}
\textup{argmin}\, S\sum_{i=1}^{k}\sum_{x\in S_i}\left \| x - u_{i} \right \|^2
\end{equation}

\section{Descriptive Statistics}

\begin{figure}[h]
    \centering
       \begin{subfigure}[t]{0.49\textwidth}
            \caption{\textbf{A) NASDAQ}}
     \includegraphics[width=\textwidth]{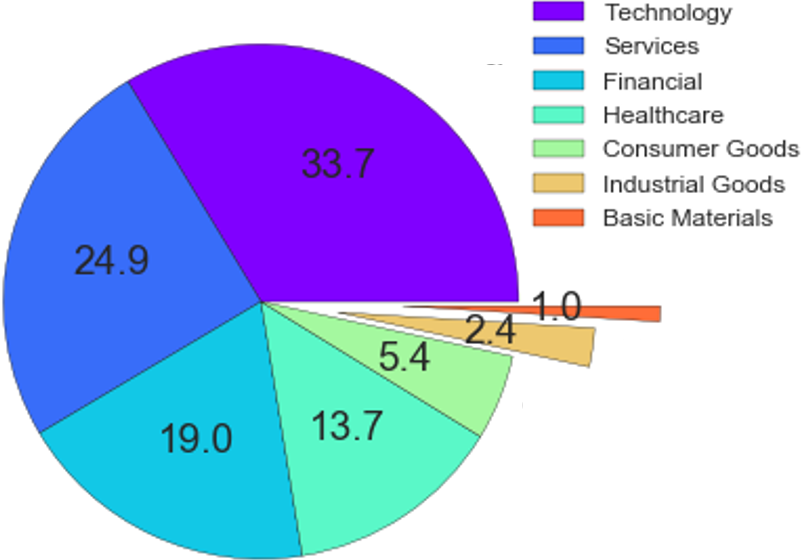}
    \end{subfigure}
     \begin{subfigure}[t]{0.49\textwidth}
          \caption{\textbf{B) NYSE}}
     \includegraphics[width=\textwidth]{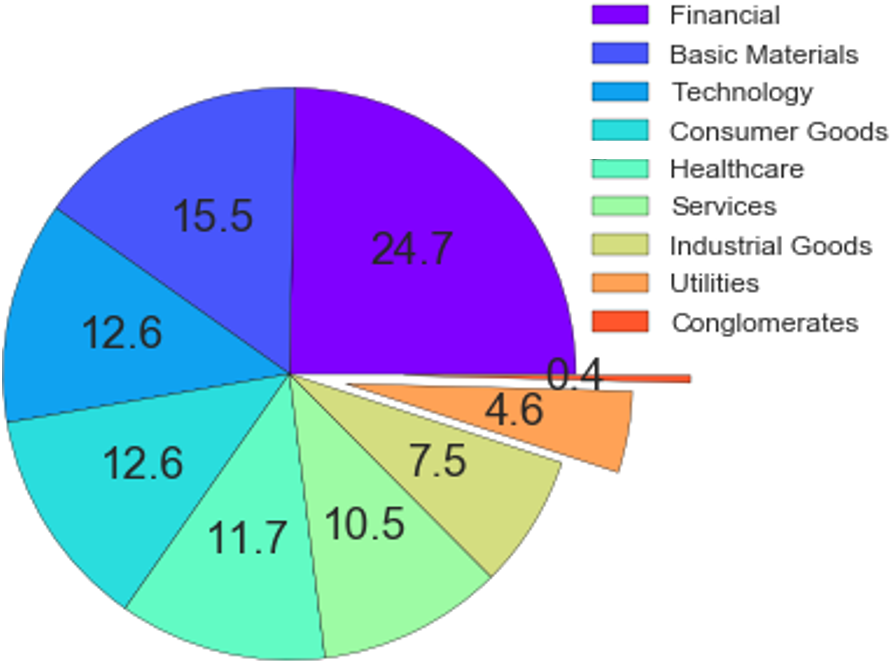}
    \end{subfigure}
    \caption{\textbf{Sector breakdown of selected companies by exchange.} Percentage breakdown by sector highlights both that the majority of large cap stocks belong to technology and financial sectors and that the NYSE has a more balanced distribution.}
    \label{pies}
\end{figure}

The following descriptive statistics provide context for our results. 
Figure \ref{pies} presents the sector breakdown of the top 250 stocks by market cap for both the NASDAQ and NYSE (i.e. 500 stocks in total). 
In the context of this work, it is interesting to note that the NASDAQ is typically characterized as being more volatile than the NYSE exchange \cite{vol2}.

\FloatBarrier

\begin{figure}[ht]
    \centering
     \begin{subfigure}[t]{0.49\textwidth}
     \caption{\textbf{A) NASDAQ}}
     \includegraphics[width=\textwidth]{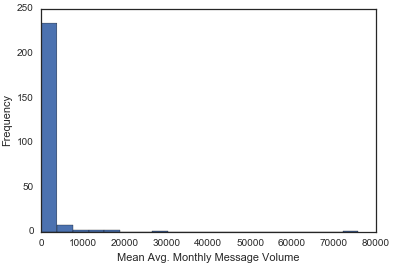}
    \end{subfigure}
     \begin{subfigure}[t]{0.49\textwidth}
     \caption{\textbf{B) NYSE}}
     \includegraphics[width=\textwidth]{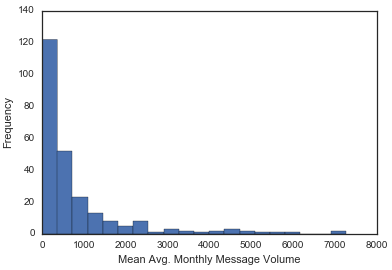}
    \end{subfigure}
    \caption{\textbf{Probability distribution of tweet volume.} There is a wide discrepancy between the NYSE and NASDAQ tweet volumes for the selected companies. 
    The NASDAQ contains several outliers (most notably, \$APPL) that skew the distribution towards the high end. 
    The NASDAQ contains a larger number and a more variable distribution of tweets, with a total of 324,239, and a standard deviation of 5,489, 
    in contrast to 212,368 and 1,241, respectively, for the NYSE.}
    \label{hist}
\end{figure}
\FloatBarrier

\begin{figure}[!h]
    \centering
     \begin{subfigure}[t]{0.49\textwidth}
     \caption{\textbf{A) NASDAQ}}
     \includegraphics[width=\textwidth]{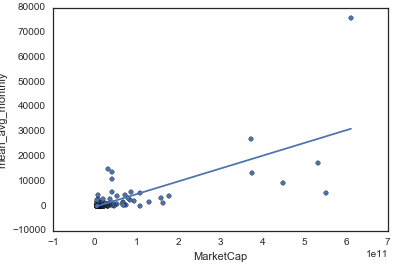}
    \end{subfigure}
     \begin{subfigure}[t]{0.49\textwidth}
     \caption{\textbf{B) NYSE}}
     \includegraphics[width=\textwidth]{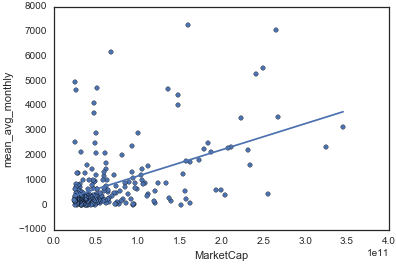}
    \end{subfigure}
    \caption{\textbf{Market cap and Tweet Volume Positive Relationship .} The NYSE and NASDAQ selected companies exhibit a 
    moderate positive correlation between market capitalization (x) and average monthly tweet volume (y).}
    \label{regression}
\end{figure}

Figures \ref{hist} and \ref{regression} compare the relationship between the volume of tweets regarding 
exchange-specific securities and the size of those securities. 
In summary, there is a strong right skew in tweet volume, which contributes to a moderately 
positive relationship between the size of a security and its interest to investors as quantified by tweet volume. 
In both exchanges, there are several notable outliers, such as \$AAPL that contains a disproportionate volume of tweets.

\begin{figure}
    \centering
    \includegraphics[width=.95\textwidth]{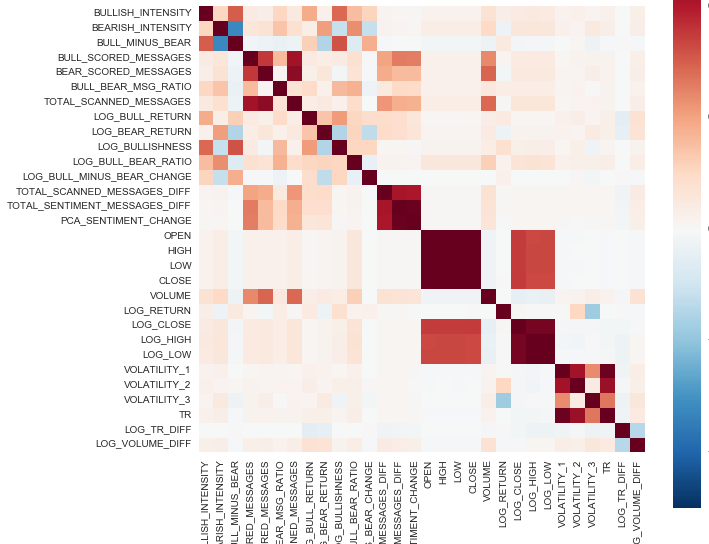}
    \caption{\textbf{NASDAQ correlation matrix.} Upper right (and low left) quadrant reveals weak to no correlation between social and financial features. The NYSE features exhibit comparable correlation.}
    \label{cor_plot}
\end{figure}

\begin{figure}[!h]
    \centering
    \begin{subfigure}[t]{0.45\textwidth}
    \centering
    \caption{\textbf{A)}}
     \includegraphics[width=\textwidth]{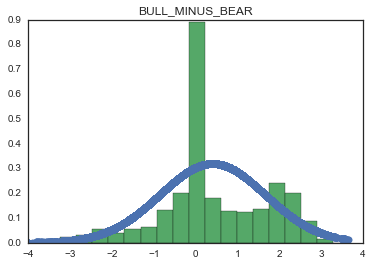}
    \end{subfigure}
    ~
     \begin{subfigure}[t]{0.45\textwidth}
     \centering
     \caption{\textbf{B)}}
     \includegraphics[width=\textwidth]{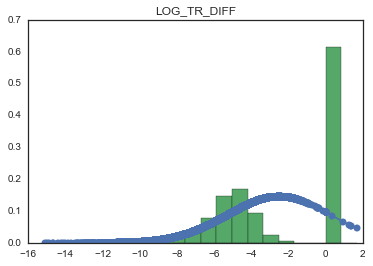}
     \end{subfigure}
    \caption{\textbf{Probability Distribution Examples with Fitted Normal Curve.} The plots are exemplary of the non-normal distributions found in both our social media (A) and financial features (B) for the NASDAQ and NYSE. }
    \label{pdfplot}
\end{figure}

In Figures \ref{cor_plot} and \ref{pdfplot}, 
we present a range of plots related to the feature behavior of the data utilized for calculating the $MI$. 
The correlation plot between all features (see tables \ref{sentvariables} and \ref{finvariables}) in Fig. \ref{cor_plot} reinforces the use of 
PCA dimensionality reduction. As the features are correlated, it is useful to inject a compression of these variables into our MI calculation. 
A correlation was found among social media features pertaining to, for instance, sentiment polarity, the volume of messages as well as with daily financial features.

The majority of features is also characterized by non-normal distributions (see Fig. \ref{pdfplot}). 
For features that are highly dependent on market capitalization (e.g., mean average monthly message volume as (see Figure \ref{hist})), 
log-normal behavior exists. Furthermore, NASDAQ and NYSE stocks exchanges exhibit comparable results.

\section{Results}
We used the outlined information surplus method~\cite{citeulike:13108056} and data from 01/01/2012 to 01/01/2016 
to determine which stocks in the NYSE and NASDAQ exhibit, on average, a significant leading information surplus. 
We then built on this method by clustering stocks and examining which configuration of variables is indicative of a high information surplus.

\subsection{Reducing Uncertainty about Volatility}

\begin{figure}[h]
    \centering
    \includegraphics[width=0.8\textwidth]{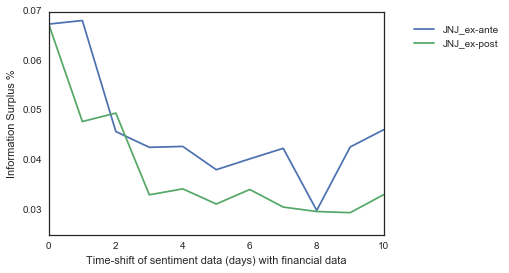}
    \caption{\textbf{Sample Information Surplus.} For each lag, the surplus is calculated as the amount of $MI$ above the base case, $MI_{t=0}$. 
    A significant \textit{ex-ante} surplus occurs where the surplus is greater than the average \textit{ex-post} surplus for the 10-day window.}
    \label{MI_sample}
\end{figure}

Experiments on the NASDAQ and NYSE were carried out in tandem, and they produced comparable results. 
We found 101 stocks from the NASDAQ that exhibit a leading information surplus when using social media as an indicator of the daily change in True Range (our measure of volatility, see Equation \ref{TRequation}). 
Of the original 250 stocks examined, 149 did not have a significant surplus, meaning that the surplus in each time lag over a 10-day 
period did not exceed the expected surplus. Thus, for example, the periods in Fig. \ref{MI_sample} where the blue \textit{ex-ante} 
series is below the average ex-post for the 
10-day period do not contain a significant leading surplus. 
All 101 stocks with a significant leading surplus, however, passed through the second validation test by performing better than 95\% of the randomly permuted data. 

For the NYSE, 91 of the original 250 stocks exhibited a significant leading surplus. In addition, only one company did not pass the second validation test. 

The full list of significant stocks from the NASDAQ and NYSE can be found in Appendix \ref{appendix}.

\subsection{Leading Surplus Indicators}

\begin{figure}[!h]
    \centering
   \begin{subfigure}[t]{0.45\textwidth}
   \centering
     \includegraphics[width=\textwidth]{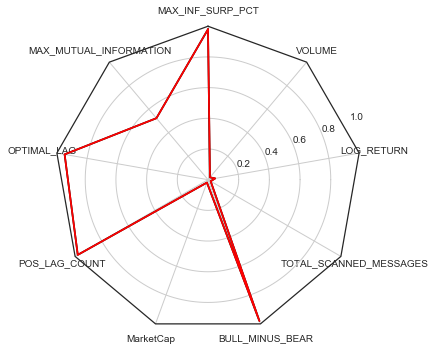}
     \end{subfigure}
     ~
    \begin{subfigure}[t]{0.45\textwidth}
    \centering
     \includegraphics[width=\textwidth]{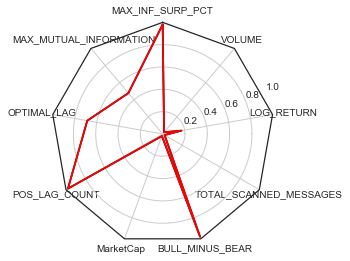}
     \end{subfigure}
    \caption{\textbf{Radar Plot of Cluster Centroid Feature Profiles.} In both the NASDAQ and NYSE, we observe a cluster that contains the majority of leading surplus lags (\texttt{POS\_LAG\_COUNT}) and maximum surplus (\texttt{MAX\_INF\_SURP\_PCT}) for $k=\left \{ 2,3,4,5,6,7 \right \}$. The important relationship represented in this cluster is the high value of \texttt{BULL\_MINUS\_BEAR}.}
    \label{radar}
\end{figure}

Stocks with a statistically significant information surplus were clustered using relevant numeric features studied (see Tables \ref{sentvariables} and \ref{finvariables}).  
The aim of the clustering is to identify the cluster feature profile of stocks with a surplus exhibited at many time lags (i.e. high \texttt{POS\_LAG\_COUNT} values) 
and the stocks with the highest surplus (i.e. high \texttt{MAX\_INF\_SURP\_PCT} values). 
The results were robust across different values of number of clusters \textit{k} tested. 
We found that moderate choices of \textit{k} all produced the noteworthy cluster profile presented in Fig. \ref{radar}, 
where a significant surplus is almost completely contained. 
The two cluster profiles presented in Fig. \ref{radar} correspond to the cluster centroid feature profiles of both the NASDAQ and NYSE which contained the highest information surplus values. 

Interestingly, \texttt{BULL\_MINUS\_BEAR} is maximized in the cluster, 
indicating that stocks with many leading surpluses during the 10-day period (\texttt{POS\_LAG\_COUNT}) 
and large significant surpluses (\texttt{MAX\_INF\_SURP\_PERC}) also have a particular ratio of bullish to bearing intensity. 
This suggests that in isolation, messages with a strong positive or negative sentiment are not as informative about future volatility; 
rather, the combination of features is important in the context of volatility or daily risk. 
Although we can speculate about the underlying factors contributing to this behavior, we note that this feature relationship seems to contradict the traditionally held assumption that \textit{negative} sentiment is more indicative of volatility. 
It may be expected that a volatile security is lead by historically negative or bearish tweets because volatility is associated with risk, 
and high risk is synonymous with negative sentiment; however, our clustering suggests otherwise: 
the polarity between all sentiment-loaded tweets is important in predicting volatility. 
Another noteworthy implication of the clustering is the absence of \texttt{VOLUME} (number of trades), 
\texttt{LOG\_RETURN}, and \texttt{TOTAL\_SCANNED\_MESSAGES} (volume of messages) in the cluster profile associated with significant leading surpluses. 
While those are features are commonly used in the literature, in this study they did not prove to be as relevant as the ratio of bullish to bearish messages.

\section{Conclusion}

Our results demonstrate that signals from social media can lead daily financial volatility in a large proportion of the 500 stocks 
examined from the NASDAQ and NYSE. A total of 101 (40\%) stocks from the NASDAQ and a further 91 stocks (36\%) from the NYSE exhibited a 
statistically significant information surplus. This was found by identifying an increase in the mutual information between social and financial time series. 

While our framework of analysis is closely aligned with earlier works~\cite{citeulike:13108056}, our results are novel. 
First, by identifying an information surplus in a large number of stocks, 
we found that social media has the capability to lead financial markets in a much larger segment of the market than in previous works, 
which only reported 12 stocks~\cite{citeulike:13108056} of significant lead-time information.

A key aim of this work was to go beyond the determination of the predictive capability of social media and attempt to determine the 
configuration of features with which this occurs.  
Our results revealed that stocks with the highest net sentiment polarity had the highest information surpluses. 
Interestingly, in contrast to other works,  stocks with a high information surplus did not require a high volume of messages and did not have a high average log-return. 
To identify this, we characterized each stock using an average representation of social and financial variables using PCA analysis, and we 
applied a clustering algorithm to inspect the group of stocks that exhibited the highest maximum information surplus.



In summary, we have challenged the notion of the efficient market hypothesis by examining the effect of the continuously 
evolving source of information embedded in social media and its effect in financial volatility. 
Using a method rooted in information theory, we have presented results that have identified a large set 
of stocks for which social media can be informative regarding financial volatility. 
By clustering stocks based on the joint feature sets of social and financial variables, our research has taken an important first step in 
characterizing the conditions under which this can be the case. 
The results indicate that social media is most informative about financial market volatility when the ratio of bullish to bearish sentiment 
is high, even when the number of messages is low. 
The reported results are reproducible via an open-source library, which allows the methodology outlined in this 
research paper to be re-used for sentiment analysis applied to finance. 

\section*{Acknowledgment}
This work was supported by PsychSignal for providing social media sentiment analytics data. We also thank Prof. Tomaso Aste for valuable comments provided.

\bibliographystyle{unsrt}
\bibliography{chapter}

\appendices

\section{Open Source Package}\label{O:S}

The functionality of the open source package\footnote{sentisignal package \url{https://github.com/jonathanmanfield/sentisignal}} enables social-financial analytics development in Python. Tools are included for the following purposes:
\begin{itemize}
    \item web scraping of historic financial records
    \item data fusion of social-financial databases
    \item measuring of information surplus 
    \item statistical significance testing
    \item cluster analysis
    \item data visualization
\end{itemize}

\section{Significant Stocks Table}
\label{appendix}
Tables \ref{tb:nasdaq} and \ref{tb:nyse} contain the NASDAQ and NYSE stocks that exhibit a significant leading surplus.
The symbols are organized alphabetically, and the columns include the maximum surplus expressed as a percent of the $MI$ 
above the baseline exhibited over the 10-day window. 
The max lag is the day when this maximum occurs (e.g. -7 $=$ one week prior). The average surplus is the expected surplus over the 10-day window. 
Each day with a significant leading surplus greater than 0 is tallied in the count column. The final column is the sector of the corresponding stock. 

\clearpage
\onecolumn

\begin{longtable}{rlccccc}
\caption{NASDAQ Significant Leading Companies}\label{tb:nasdaq}\\
  \hline
 & Symbol & Max\_Surplus & Max\_Lag & Avg\_Surplus & Count & Sector \\ 
  \hline
1 & AAL & 5.98 &  -7 & 0.60 &   1 & Transportation \\ 
  2 & ACGL & 16.46 & -10 & 5.86 &   6 & Finance \\ 
  3 & ACWI & 76.02 & -10 & 8.78 &   2 & n/a \\ 
  4 & ADBE & 23.44 &  -1 & 2.34 &   1 & Technology \\ 
  5 & AFSI & 0.36 &  -1 & 0.04 &   1 & Finance \\ 
  6 & AKAM & 15.02 &  -1 & 1.50 &   1 & Miscellaneous \\ 
  7 & ANSS & 2.54 &  -1 & 0.25 &   1 & Technology \\ 
  8 & ASML & 20.80 &  -1 & 2.08 &   1 & Technology \\ 
  9 & BBBY & 30.15 &  -1 & 3.01 &   1 & Consumer Services \\ 
  10 & BUFF & 8.83 &  -3 & 0.88 &   1 & Consumer Non-Durables \\ 
  11 & CA & 149.77 &  -7 & 44.00 &   7 & Technology \\ 
  12 & CASY & 9.28 &  -1 & 0.93 &   1 & Consumer Durables \\ 
  13 & CDNS & 2.66 & -10 & 0.34 &   2 & Technology \\ 
  14 & CG & 17.93 &  -2 & 4.07 &   3 & Finance \\ 
  15 & CINF & 22.04 &  -7 & 3.01 &   2 & Finance \\ 
  16 & CMCSA & 6.33 &  -1 & 0.63 &   1 & Consumer Services \\ 
  17 & CME & 16.55 &  -1 & 1.65 &   1 & Finance \\ 
  18 & COST & 16.93 &  -4 & 2.84 &   3 & Consumer Services \\ 
  19 & CSAL & 21.08 & -10 & 2.14 &   2 & Consumer Services \\ 
  20 & CSGP & 22.04 &  -1 & 3.57 &   4 & Miscellaneous \\ 
  21 & CTAS & 28.09 &  -1 & 3.81 &   2 & Consumer Non-Durables \\ 
  22 & CTXS & 2.61 &  -1 & 0.26 &   1 & Technology \\ 
  23 & DISCB & 9.46 &  -3 & 1.17 &   2 & Consumer Services \\ 
  24 & DISCK & 0.72 &  -2 & 0.07 &   1 & Consumer Services \\ 
  25 & DOX & 52.25 &  -7 & 9.85 &   5 & Technology \\ 
  26 & EA & 16.98 &  -1 & 1.70 &   1 & Technology \\ 
  27 & ERIE & 3.85 &  -9 & 0.38 &   1 & Finance \\ 
  28 & EWBC & 34.58 &  -3 & 12.88 &   6 & Finance \\ 
  29 & FANG & 3.40 &  -8 & 0.59 &   2 & Energy \\ 
  30 & FB & 18.25 &  -1 & 1.82 &   1 & Technology \\ 
  31 & FFIV & 22.07 &  -1 & 2.21 &   1 & Technology \\ 
  32 & FISV & 4.59 & -10 & 0.46 &   1 & Technology \\ 
  33 & FITB & 26.25 &  -6 & 5.68 &   3 & Finance \\ 
  34 & FOXA & 13.71 &  -1 & 1.37 &   1 & Consumer Services \\ 
  35 & FTNT & 20.34 &  -1 & 2.03 &   1 & Technology \\ 
  36 & GLPI & 5.94 &  -8 & 1.51 &   3 & Consumer Services \\ 
  37 & GNTX & 1.81 &  -1 & 0.18 &   1 & Capital Goods \\ 
  38 & HAS & 30.73 &  -7 & 14.56 &   6 & Consumer Non-Durables \\ 
  39 & HDS & 11.63 &  -1 & 1.33 &   2 & Consumer Services \\ 
  40 & HOLX & 1.81 &  -1 & 0.18 &   1 & Health Care \\ 
  41 & HSIC & 20.18 &  -1 & 3.47 &   2 & Health Care \\ 
  42 & IBKR & 2.55 &  -1 & 0.26 &   1 & Finance \\ 
  43 & INFO & 108.15 &  -7 & 32.00 &   4 & Technology \\ 
  44 & INTU & 4.26 &  -5 & 1.01 &   4 & Technology \\ 
  45 & JBHT & 12.73 &  -1 & 1.27 &   1 & Transportation \\ 
  46 & JD & 38.70 &  -5 & 12.18 &   6 & Consumer Services \\ 
  47 & JKHY & 20.47 &  -8 & 4.89 &   6 & Technology \\ 
  48 & KLAC & 5.28 &  -1 & 0.53 &   1 & Capital Goods \\ 
  49 & LBRDA & 14.82 &  -1 & 1.57 &   2 & Consumer Services \\ 
  50 & LBTYA & 8.17 &  -1 & 0.82 &   1 & Consumer Services \\ 
  51 & LBTYB & 40.41 &  -7 & 12.91 &   7 & Consumer Services \\ 
  52 & LBTYK & 19.43 &  -7 & 3.04 &   2 & Consumer Services \\ 
  53 & LILA & 35.42 & -10 & 3.54 &   1 & Consumer Services \\ 
  54 & LILAK & 39.67 &  -7 & 8.16 &   3 & Consumer Services \\ 
  55 & LVNTA & 26.84 &  -1 & 4.91 &   4 & Consumer Services \\ 
  56 & MAR & 27.35 &  -7 & 5.02 &   2 & Consumer Services \\ 
  57 & MIDD & 11.30 &  -1 & 1.50 &   2 & Technology \\ 
  58 & MNST & 13.12 &  -2 & 1.31 &   1 & Consumer Non-Durables \\ 
  59 & MSCC & 25.80 &  -7 & 5.98 &   5 & Technology \\ 
  60 & NTAP & 26.26 &  -1 & 2.63 &   1 & Technology \\ 
  61 & NTRS & 11.51 &  -1 & 1.15 &   1 & Finance \\ 
  62 & NWS & 44.64 &  -4 & 11.26 &   5 & Consumer Services \\ 
  63 & NWSA & 3.65 &  -1 & 0.36 &   1 & Consumer Services \\ 
  64 & ON & 24.65 &  -2 & 4.50 &   2 & Technology \\ 
  65 & ORLY & 21.53 &  -1 & 2.15 &   1 & Consumer Services \\ 
  66 & PACW & 4.40 &  -8 & 0.44 &   1 & Finance \\ 
  67 & PAYX & 21.76 &  -1 & 2.18 &   1 & Consumer Services \\ 
  68 & PBCT & 17.27 &  -4 & 3.10 &   3 & Finance \\ 
  69 & PDCO & 34.06 &  -1 & 3.93 &   2 & Health Care \\ 
  70 & PPC & 48.37 &  -5 & 18.08 &   5 & Consumer Non-Durables \\ 
  71 & PYPL & 0.15 &  -9 & 0.01 &   1 & Miscellaneous \\ 
  72 & QCOM & 21.89 &  -1 & 2.19 &   1 & Technology \\ 
  73 & QGEN & 38.94 &  -1 & 7.80 &   3 & Health Care \\ 
  74 & QVCA & 17.89 &  -4 & 6.90 &   5 & Consumer Services \\ 
  75 & RYAAY & 4.68 &  -6 & 0.55 &   2 & Transportation \\ 
  76 & SABR & 35.74 &  -1 & 3.57 &   1 & Technology \\ 
  77 & SBAC & 17.38 &  -1 & 3.00 &   2 & Consumer Services \\ 
  78 & SCZ & 42.50 &  -7 & 8.77 &   5 & n/a \\ 
  79 & SIVB & 30.48 &  -6 & 7.09 &   4 & Finance \\ 
  80 & SNH & 41.11 &  -5 & 4.50 &   2 & Consumer Services \\ 
  81 & SNPS & 13.11 &  -1 & 1.31 &   1 & Technology \\ 
  82 & SSNC & 18.27 &  -1 & 2.93 &   3 & Technology \\ 
  83 & STLD & 11.49 &  -6 & 1.15 &   1 & Basic Industries \\ 
  84 & SYMC & 1.92 &  -1 & 0.19 &   1 & Technology \\ 
  85 & TEAM & 219.09 & -10 & 76.32 &   4 & Technology \\ 
  86 & TFSL & 2.09 &  -1 & 0.21 &   1 & Finance \\ 
  87 & TROW & 2.61 &  -8 & 0.52 &   3 & Finance \\ 
  88 & TSCO & 0.26 &  -1 & 0.03 &   1 & Consumer Services \\ 
  89 & UHAL & 17.33 &  -9 & 1.73 &   1 & Consumer Services \\ 
  90 & ULTI & 24.29 &  -1 & 3.35 &   3 & Technology \\ 
  91 & VCIT & 69.82 &  -8 & 13.20 &   4 & n/a \\ 
  92 & VIP & 9.70 &  -4 & 0.97 &   1 & Public Utilities \\ 
  93 & VRSK & 7.91 &  -6 & 0.79 &   1 & Technology \\ 
  94 & VRSN & 13.25 &  -1 & 1.32 &   1 & Technology \\ 
  95 & VXUS & 11.17 &  -2 & 1.71 &   3 & n/a \\ 
  96 & WDC & 5.33 &  -1 & 0.53 &   1 & Technology \\ 
  97 & WFM & 53.08 &  -1 & 5.31 &   1 & Consumer Services \\ 
  98 & WOOF & 5.13 &  -8 & 0.54 &   2 & Consumer Non-Durables \\ 
  99 & Z & 43.88 &  -1 & 17.12 &   9 & Miscellaneous \\ 
  100 & ZG & 85.96 &  -9 & 18.36 &   4 & Miscellaneous \\ 
  101 & ZION & 31.92 &  -1 & 3.66 &   3 & Finance \\ 
   \hline
\hline
\end{longtable}

\begin{longtable}{rlccccc}
\caption{NYSE Significant Leading Companies}\label{tb:nyse} \\
  \hline
 & Symbol & Max\_Surplus & Max\_Lag & Avg\_Surplus & Count & Sector \\ 
  \hline
1 & ABB & 52.00 &  -4 & 20.95 &   8 & Consumer Durables \\ 
  2 & ABEV & 0.71 &  -1 & 0.07 &   1 & Consumer Non-Durables \\ 
  3 & ABT & 161.83 & -10 & 29.49 &   4 & Health Care \\ 
  4 & ADM & 36.02 & -10 & 3.60 &   1 & Consumer Non-Durables \\ 
  5 & AEP & 44.91 &  -1 & 5.31 &   2 & Public Utilities \\ 
  6 & AFL & 17.21 &  -1 & 1.72 &   1 & Finance \\ 
  7 & AGN & 17.41 &  -6 & 2.11 &   2 & Health Care \\ 
  8 & ALL & 13.99 &  -1 & 2.37 &   2 & Finance \\ 
  9 & AON & 1.84 &  -1 & 0.18 &   1 & Finance \\ 
  10 & APC & 3.40 &  -5 & 0.34 &   1 & Energy \\ 
  11 & APD & 5.80 &  -1 & 0.58 &   1 & Basic Industries \\ 
  12 & AVB & 4.89 &  -4 & 1.09 &   3 & Consumer Services \\ 
  13 & BA & 13.08 &  -7 & 1.85 &   2 & Capital Goods \\ 
  14 & BAC & 8.54 &  -1 & 0.85 &   1 & Finance \\ 
  15 & BAM & 3.24 &  -8 & 0.32 &   1 & Consumer Services \\ 
  16 & BAX & 27.76 &  -1 & 2.78 &   1 & Health Care \\ 
  17 & BBD & 3.07 &  -5 & 0.31 &   1 & Finance \\ 
  18 & BBT & 12.70 &  -1 & 1.27 &   1 & Finance \\ 
  19 & BBVA & 0.90 &  -6 & 0.09 &   1 & Finance \\ 
  20 & BMO & 6.82 &  -8 & 0.83 &   2 & Finance \\ 
  21 & BNS & 20.90 &  -6 & 2.09 &   1 & Finance \\ 
  22 & BSBR & 8.60 &  -1 & 0.86 &   1 & Finance \\ 
  23 & BT & 27.21 &  -1 & 2.88 &   2 & Public Utilities \\ 
  24 & CAJ & 16.87 &  -1 & 1.69 &   1 & Miscellaneous \\ 
  25 & CAT & 139.16 &  -6 & 41.02 &   6 & Capital Goods \\ 
  26 & CHA & 45.38 &  -1 & 20.53 &   5 & Public Utilities \\ 
  27 & COP & 1.84 &  -7 & 0.18 &   1 & Energy \\ 
  28 & CRH & 11.57 &  -8 & 1.16 &   1 & Capital Goods \\ 
  29 & CS & 10.87 &  -8 & 1.24 &   2 & Finance \\ 
  30 & CUK & 2.56 &  -3 & 0.26 &   1 & Consumer Services \\ 
  31 & DAL & 9.74 & -10 & 3.21 &   4 & Transportation \\ 
  32 & DCM & 8.48 &  -6 & 0.87 &   2 & Technology \\ 
  33 & DHR & 2.81 &  -1 & 0.28 &   1 & Capital Goods \\ 
  34 & DIS & 19.97 &  -4 & 2.47 &   3 & Consumer Services \\ 
  35 & DOW & 267.29 &  -6 & 87.33 &  10 & Basic Industries \\ 
  36 & DUK & 7.83 &  -1 & 0.78 &   1 & Public Utilities \\ 
  37 & EMR & 17.66 &  -7 & 6.15 &   7 & Energy \\ 
  38 & EOG & 33.09 &  -9 & 5.21 &   3 & Energy \\ 
  39 & EPD & 16.03 &  -4 & 4.25 &   4 & Public Utilities \\ 
  40 & FMX & 37.36 &  -7 & 4.34 &   2 & Consumer Non-Durables \\ 
  41 & GGP & 17.05 & -10 & 2.10 &   2 & Consumer Services \\ 
  42 & HMC & 3.52 &  -2 & 0.35 &   1 & Capital Goods \\ 
  43 & HPE & 44.37 &  -7 & 8.20 &   3 & Technology \\ 
  44 & HPQ & 0.30 &  -1 & 0.03 &   1 & Technology \\ 
  45 & HUM & 76.89 &  -7 & 39.40 &   8 & Health Care \\ 
  46 & IBM & 26.29 & -10 & 7.16 &   4 & Technology \\ 
  47 & ING & 11.96 &  -8 & 2.73 &   3 & Finance \\ 
  48 & JNJ & 12.29 &  -1 & 1.23 &   1 & Health Care \\ 
  49 & JPM & 2.67 &  -1 & 0.27 &   1 & Finance \\ 
  50 & KEP & 48.05 &  -9 & 11.22 &   6 & Public Utilities \\ 
  51 & KO & 2.83 &  -1 & 0.28 &   1 & Consumer Non-Durables \\ 
  52 & LFC & 10.22 &  -5 & 2.19 &   4 & Finance \\ 
  53 & LYG & 11.33 &  -1 & 1.13 &   1 & Finance \\ 
  54 & MFC & 0.03 &  -1 & 0.00 &   1 & Finance \\ 
  55 & MFG & 4.11 &  -2 & 0.41 &   1 & Finance \\ 
  56 & MMC & 22.17 &  -3 & 4.05 &   4 & Finance \\ 
  57 & MMM & 5.85 &  -1 & 0.59 &   1 & Health Care \\ 
  58 & MS & 7.49 &  -1 & 0.75 &   1 & Finance \\ 
  59 & MTU & 35.42 &  -2 & 8.86 &   6 & Finance \\ 
  60 & ORAN & 28.61 & -10 & 11.02 &   7 & Public Utilities \\ 
  61 & PCG & 9.17 & -10 & 0.92 &   1 & Public Utilities \\ 
  62 & PLD & 1.27 &  -9 & 0.13 &   1 & Consumer Services \\ 
  63 & PPG & 26.29 &  -5 & 3.91 &   2 & Basic Industries \\ 
  64 & PRU & 13.98 &  -5 & 4.81 &   5 & Finance \\ 
  65 & PSA & 16.23 &  -8 & 3.22 &   4 & Consumer Services \\ 
  66 & PTR & 17.52 &  -8 & 3.13 &   2 & Energy \\ 
  67 & PUK & 37.53 &  -7 & 12.24 &   5 & Finance \\ 
  68 & RAI & 0.71 &  -1 & 0.07 &   1 & Consumer Non-Durables \\ 
  69 & RIO & 7.35 &  -9 & 0.74 &   1 & Basic Industries \\ 
  70 & RTN & 15.74 &  -1 & 1.57 &   1 & Capital Goods \\ 
  71 & SAN & 47.81 & -10 & 13.18 &   5 & Finance \\ 
  72 & SMFG & 124.01 &  -3 & 44.31 &   6 & Finance \\ 
  73 & SNY & 25.29 &  -9 & 5.69 &   4 & Health Care \\ 
  74 & SRE & 49.66 &  -6 & 20.94 &   7 & Public Utilities \\ 
  75 & T & 13.15 &  -1 & 1.31 &   1 & Public Utilities \\ 
  76 & TD & 6.17 &  -1 & 0.62 &   1 & Finance \\ 
  77 & TEF & 12.43 &  -3 & 1.51 &   2 & Public Utilities \\ 
  78 & TJX & 4.62 &  -1 & 0.46 &   1 & Consumer Services \\ 
  79 & TLK & 28.79 &  -8 & 5.21 &   2 & Public Utilities \\ 
  80 & TM & 28.68 & -10 & 4.72 &   3 & Capital Goods \\ 
  81 & TOT & 9.79 & -10 & 0.98 &   1 & Energy \\ 
  82 & TRP & 5.26 &  -1 & 0.53 &   1 & Public Utilities \\ 
  83 & TRV & 54.89 &  -5 & 7.85 &   2 & Finance \\ 
  84 & TWX & 3.11 &  -1 & 0.31 &   1 & Consumer Services \\ 
  85 & UBS & 31.27 &  -1 & 4.70 &   3 & Finance \\ 
  86 & UN & 12.38 &  -8 & 1.96 &   2 & Basic Industries \\ 
  87 & VIV & 7.40 &  -9 & 0.84 &   2 & Public Utilities \\ 
  88 & VMW & 10.31 &  -1 & 1.03 &   1 & Technology \\ 
  89 & VTR & 12.66 &  -9 & 2.66 &   3 & Consumer Services \\ 
  90 & WBK & 37.99 & -10 & 8.98 &   4 & Finance \\ 
  91 & WMT & 5.26 &  -1 & 0.53 &   1 & Consumer Services \\ 
   \hline
\hline
\end{longtable}

\end{document}